\documentclass[aps,pre,twocolumn,showpacs,superscriptaddress,groupedaddress, floatfix]{revtex4-1}  
\usepackage{graphicx}  
\usepackage{dcolumn}   
\usepackage{bm}        
\usepackage{amssymb, amsmath}   
\usepackage{multirow} 
\usepackage{color}

\begin{document}

\title{Slow poisoning and destruction of networks: Edge proximity  and its implications for biological and infrastructure networks}
                        
\date{\today}
\author{Soumya Jyoti Banerjee}
\affiliation{Bose Institute, 93/1 Acharya Prafulla Chandra Roy Road, Kolkata 700 009 India}
\author{Saptarshi Sinha}
\affiliation{Bose Institute, 93/1 Acharya Prafulla Chandra Roy Road, Kolkata 700 009 India}
\author{Soumen Roy}
\email{soumen@jcbose.ac.in}
\affiliation{Bose Institute, 93/1 Acharya Prafulla Chandra Roy Road, Kolkata 700 009 India}

\begin{abstract}
We propose a network metric, edge proximity, ${\cal P}_e$, which demonstrates the importance of specific edges in a network, hitherto not captured by existing network metrics. The effects of removing edges with high ${\cal P}_e$ might initially seem inconspicuous but are eventually shown to be very harmful for networks. Compared to existing strategies, the removal of  edges by ${\cal P}_e$ leads to a remarkable increase in the diameter and average shortest path length in undirected real and random networks till the first disconnection and well beyond. ${\cal P}_e$ can be consistently used to rupture the network into two nearly equal  parts, thus presenting a very potent strategy to greatly harm a network. Targeting by ${\cal P}_e$ causes notable efficiency loss in U.S. and European power grid networks.  ${\cal P}_e$ identifies proteins with essential cellular functions in protein-protein interaction networks. It pinpoints regulatory neural connections and important portions of the neural and brain networks, respectively. Energy flow interactions identified by ${\cal P}_e$ form the backbone of long food web chains. Finally,  we scrutinize the potential of ${\cal P}_e$ in edge controllability dynamics of directed networks.
\end{abstract}
\maketitle
Considerable research has been done on the importance of various metrics in  complex networks~\cite{barabasi-rmp, Mark-book,epl091,pre091}. The importance of nodes and remarkable effects of their targeted removal using  various network metrics like betweenness and degree is now 
well known. In comparison, the role of edges has received less attention~\cite{Newman-bet,Girvan-bet, influ_edge}. In this context, one may ask if it is possible to prognose situations where not even a single node is pruned from the network for a long time and yet tremendous damage is inflicted on it by selective targeting of specific edges.  Indeed, as we show here, such a process might initially appear inconspicuous or even deceptively innocuous. Using a simple metric, which we call {\em edge proximity},  we are able to identify specific edges whose removal can slowly poison networks and silently wreak havoc in them. Furthermore, we show that  ${\cal P}_e$ can be used to design strategies to {\em consistently} rupture  networks into two nearly equal parts. Thus, this  could eventually be far more destructive than currently available strategies for targeting network edges including those where rapid disconnection can be achieved due to detachment of small subgraphs from the parent network.   

The most well-known edge-based measure, {\em edge betweenness}, ${\cal B}_e$, attempts to capture the frequency of an edge lying on the shortest paths between all pairs of vertices in a network~\cite{Newman-bet,Girvan-bet}. Edges with the highest  ${\cal B}_e$ are most likely to lie between subgraphs, rather than inside them. Thus, targeting by node or edge betweenness ensures rapid disconnection of networks by a small number of deletions~\cite{pre101,ssb121}. 

Herein, we introduce a new edge-based 
network metric, edge proximity, ${\cal P}_e$. The ${\cal P}_e$  of an edge, $e\in{\cal E}$, is the inverse of the sum of its shortest distance $d (e,f)$, with every other edge, $f\in{\cal E}$, in a connected network, $\cal G (V,E)$.  
$\cal V and \cal E$ denote the set of nodes and edges, respectively, in $\cal G$. $\cal N = |V|$ and $\cal M = |E|$ are the total number of nodes and edges in $\cal G$, respectively. ${\cal P}_e$ lends clues as to how close each edge is to every other edge in ${\cal G}$ through the shortest paths between them. Thus, for $e\in{\cal E}$, 

\begin{equation}
{\cal P}_e = \frac{{\cal M}-1}{\displaystyle \sum_{f\in{\cal E}} d(e,f)}
\label{edge-proximity}
\end{equation} 

The {\em Average shortest path length, $\cal L_G$,}  is the average of all the shortest path lengths between any pair of nodes in ${ \cal G}$ and is defined as,

\begin{equation} 
{\cal L_G} = \frac{1}{{\cal N(N}-1)} \sum_{s,t \in {\cal V}; ~s \ne t}{d(s, t)}
\label{eq:APL}
\end{equation}
The {\em Diameter} of  ${\cal G}$ is defined as, 
\begin{equation} 
{\cal D} = max({d(s, t))},~ {\forall~s,t \in {\cal V}; ~s \ne t};  
\label{eq:Diam}
\end{equation}

$d(s,t)$ being the shortest path from $s$ to $t$.  From the definition of $\cal L_G$ and $\cal D$ it is clear that $\cal L_G$ and $\cal D$ become infinite when $\cal G$ becomes disconnected. 

When edges are targeted by  ${\cal B}_e$, the damage done to the network in the form of increases in $\cal L_G$ and $\cal D$ might initially seem to be greater. However, we observe here that ${\cal P}_e$ helps in identifying those crucial edges of  undirected networks whose deletion ensures the highest increase in $\cal L_G$ and $\cal D$  in ${\cal G}$ (or its largest connected component, $\cal G'$) compared to other methods of edge deletion.  This appears to be true for both the first disconnection and well beyond. In fact, when targeting by  ${\cal P}_e$ no node is disconnected from the network for a very long time.  Here, we study the effect of various edge deletion strategies on real-world undirected networks like the E.U. and U.S. power grid network (PGNs) and the protein-protein interaction networks (PPINs) of  {\it S. cerevisiae} and {\it E. coli} till the first disconnection. We also scrutinize the effect of these strategies,  long after the first disconnection (till only about $30\%$ of the edges remain in $\cal G'$), on various models like the Erd\"os-R\'enyi (ER)~\cite{er}, Barabasi-Albert (BA)~\cite{ba}, and small-world (SW)~\cite{sw,sw-dis} networks. We also study a variety of directed biological networks, namely, the macaque brain network, the {\it C. elegans} neural network, and a number of food webs. We find that in each case ${\cal P}_e$ successfully provides meaningful biological information. 

There have been lots of studies on disconnection of networks by malicious targeting. However, it is obvious that significant damage would be caused to the network when each disconnection causes the network to rupture into two nearly equal parts rather than having a small chunk disconnected from it. We demonstrate that ${\cal P}_e$ can be remarkably successful in consistently achieving this, compared to other methods of edge removal. 

An edge with higher ${\cal P}_e$ should possess the potential to reach many other edges in directed networks. 
Of late, there has been considerable research on node controllability in networks~\cite{Barabasi-control, SJB-control}.  
Switchboard dynamics (SBD) of edges has also been recently proposed to study edge controllability of directed networks and to identify the minimal set of driven edges~\cite{Nepusz-control}. We conclude by examining the natural potential that ${\cal P}_e$ possesses in the context of controllability of directed networks.

\begin{figure}[http]%
\includegraphics[width=\columnwidth, height = 3.5cm]{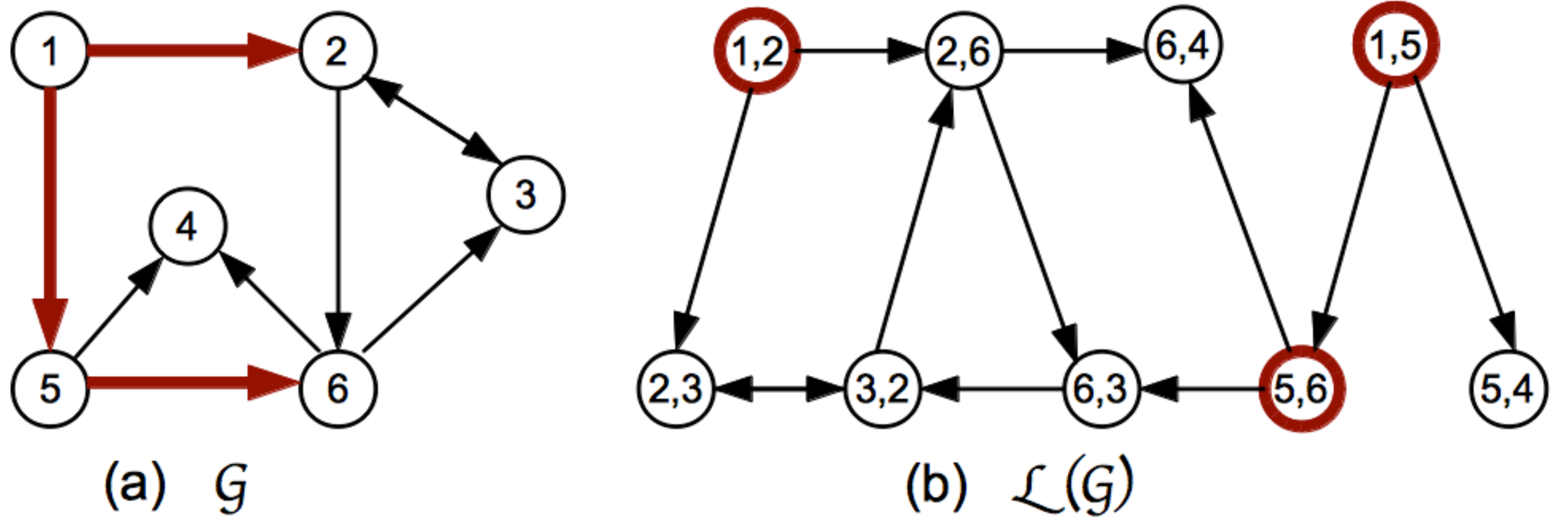}
\caption{Construction of a directed line graph~\cite{Nepusz-control}, $\cal L(G)$, from a directed graph, $\cal G$, to calculate the value of ${\cal P}_e$ for all edges. Edges in ${\cal G}$ (nodes in $\cal L(G)$) shown in {\bf {{bold}}} depict a set of driven edges found by applying the maximum matching algorithm in $\cal L(G)$.}
\label{Fig:toy}
\end{figure}

Figure~\ref{Fig:toy}  uses a toy example to illustrate  the construction of a line graph $\cal L (G)$ from $\cal G$.  The first step for creating $\cal L(G)$ is that every node in $\cal L(G)$  represents an edge in the original graph $\cal G$.  An edge is drawn between any two nodes in $\cal L(G)$ if the corresponding  edges in $\cal G$ share a common node. For directed graphs, an edge in $\cal L(G)$ represents a directed path of length $2$ in $\cal G$. Each node of  $\cal L (G)$ is an edge of $\cal G$. Thus,  ${\cal P}_e, e \in \cal E$ can be obtained from closeness centrality of corresponding node in $\cal L (G)$. 

As is well known, the computational complexity of node closeness is ${\cal O ~(|V|}~log~{\cal |V| + |E|})$. 
The number of nodes in $\cal L(G)$ is $\cal |E|$. The number of two-path lengths in $\cal G$ is the number of edges in $\cal L(G)$. Therefore the computational complexity for ${\cal P}_e$ is ${\cal O ~(|E|}~log~{\cal |E|  + |E|}^{2})$ and can be computed easily. All networks were analyzed here using NetworkX~\cite{netx}. 

The very same edge-based metric, $\cal X (G)$, can be modelled with nodes involved in definition, as ${\cal X}_1{\cal (G)}$,  or without nodes as ${\cal X}_2 {\cal (G)}$.  $|{\cal X}_1{\cal (G)} \sim {\cal X}_2{\cal (G)}|$ might not be very significant, especially when ${\cal G}$ is large.  The computation time for ${\cal X}_2{\cal (G)}$ might be slightly longer than that for ${\cal X}_1{\cal (G)}$.

We investigate a number of edge deletion strategies which affect $\cal L_G$ and $\cal D$ 
in ${\cal G}$. The strategies adopted here consist of {\em independently} deleting successive edges: (i) with $max({\cal P}_e)$, (ii) with $max ({\cal B}_e)$, (iii) connected to the node of the highest degree in ${\cal G}$ $[max({\cal K}_e)]$, and,  (iv) purely at random (${\cal R}_e$). 
To illustrate further, we construct four identical copies of ${\cal G (V,E)}$; $\{{\cal G}_i ({\cal V}_i, {\cal E}_i)\},i\in\{1, 2, 3, 4\} $. We then remove  the edge with $max ({\cal P}_{e})$,  $max({\cal B}_e)$, $max({\cal K}_e)$, $e \in {\cal E}_i, i\in\{1,2,3\}$, from 
${\cal G}_1$,   ${\cal G}_2$, ${\cal G}_3$, respectively. In case there is more than one edge with  $max ({\cal P}_{e})$, $max({\cal B}_{e})$, and $max({\cal K}_{e})$, we randomly choose one among them. We recalculate the values of ${\cal P}_{e}$,  ${\cal B}_{e}$ and ${\cal K}_{e}$, $e  \in {\cal E}_i, i\in \{1, 2, 3\}$; for ${\cal G}_1$, ${\cal G}_2$, and ${\cal G}_3$, respectively. For all real-world networks studied here, we repeat this removal and recalculation process, until the first node is disconnected. In ${\cal G}_4$  edges are always deleted randomly.

\begin{figure}[htb]%
\vspace{-0.15in}%
\includegraphics[width=0.5\columnwidth, height=5cm]{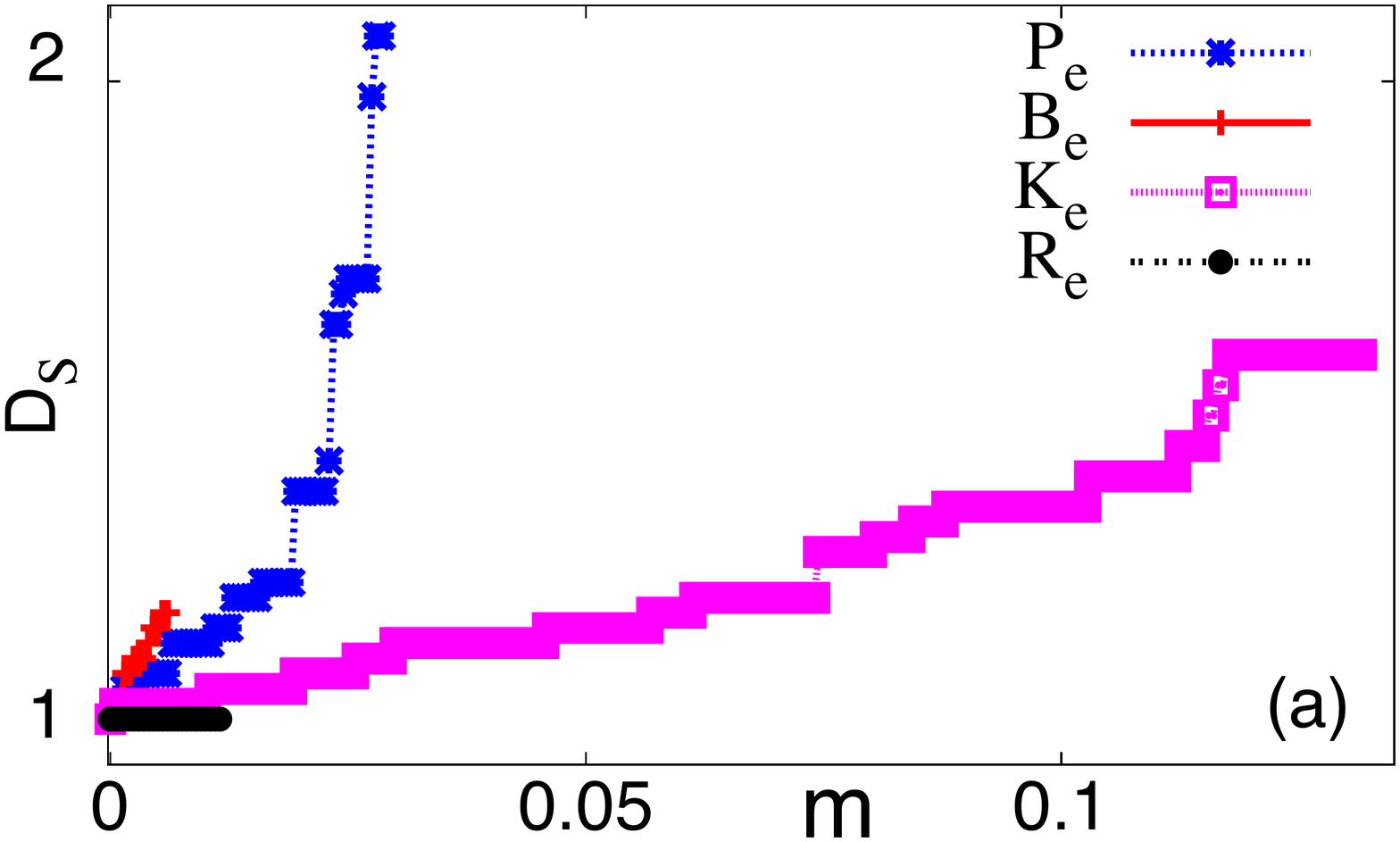}%
\vspace{-0.35in}%
\includegraphics[width=0.5\columnwidth, height=5cm]{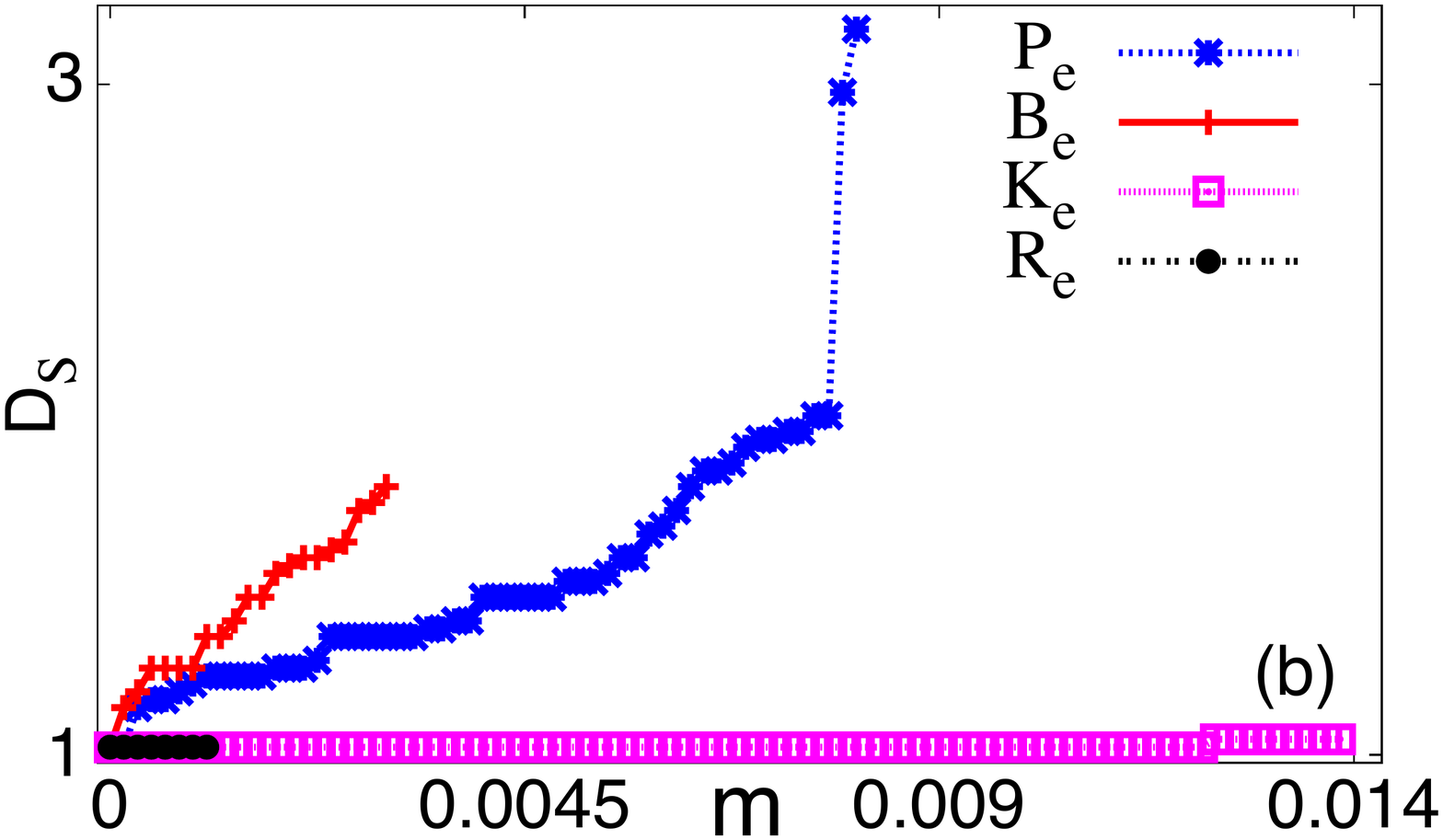}
\vspace{-0.2in}%
\includegraphics[width=0.5\columnwidth, height=5cm]{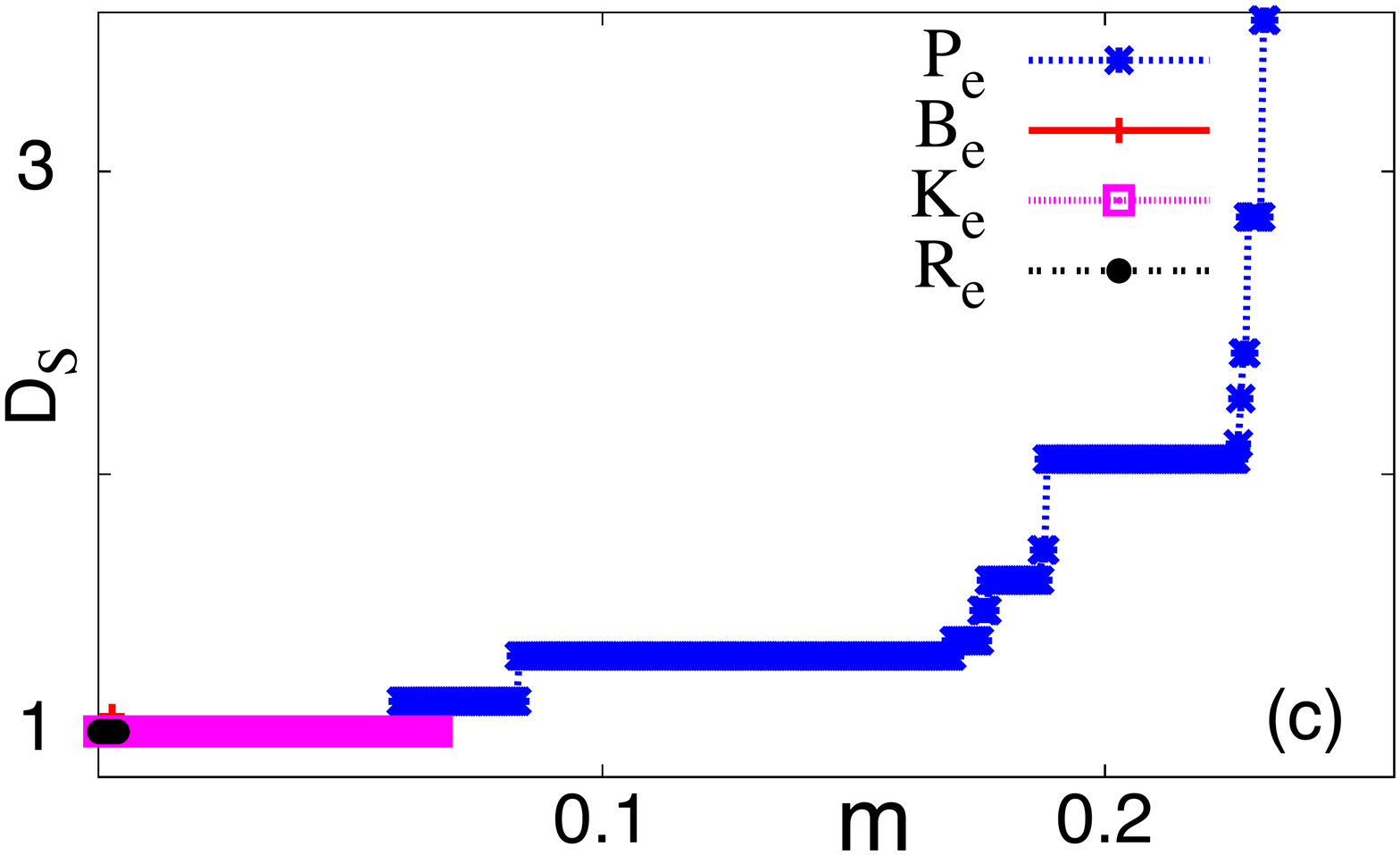}%
\includegraphics[width=0.5\columnwidth, height=5cm]{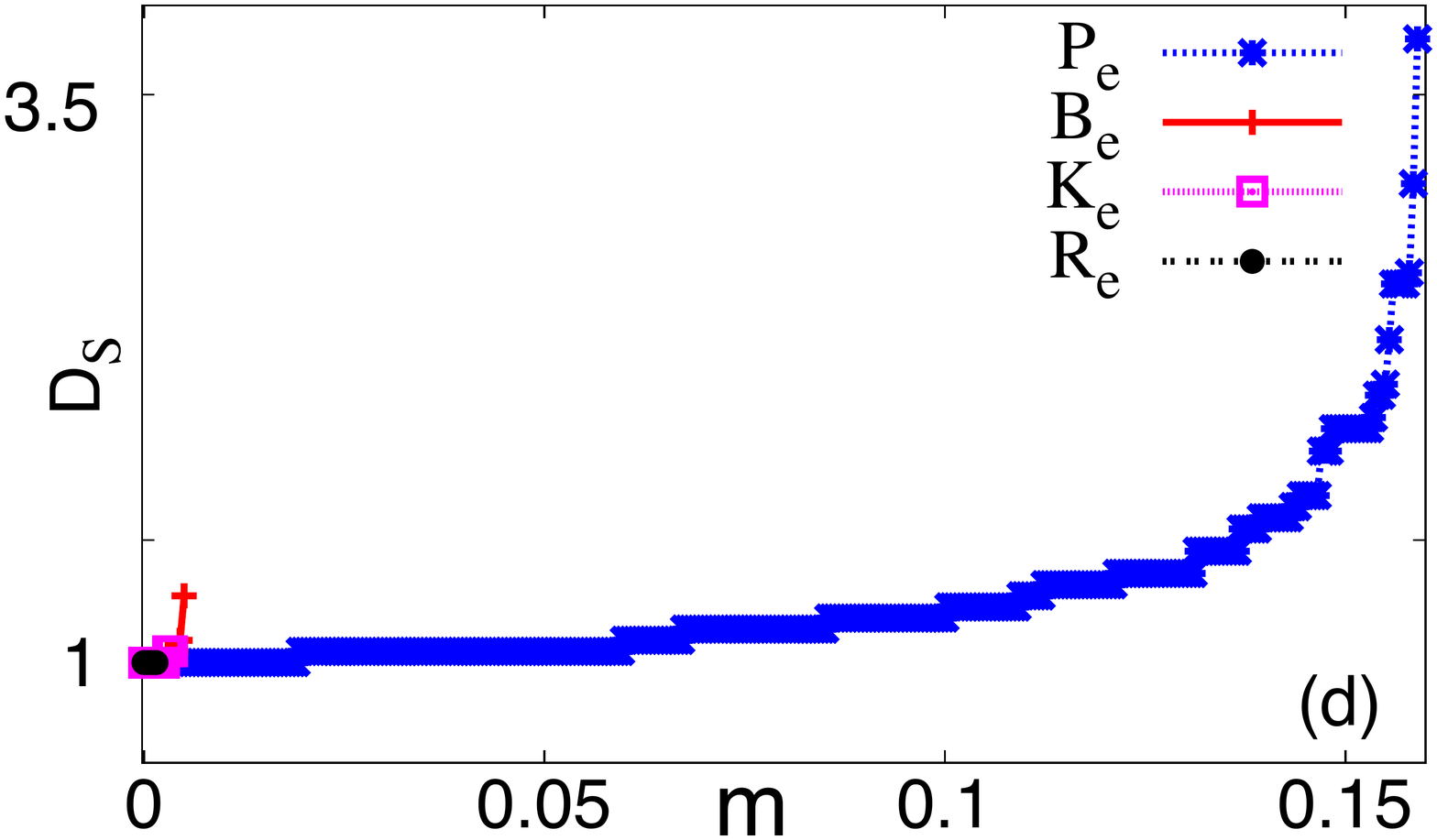}%
\vspace{-0.15in}%
\caption{${\cal D_S =  D/D}_0$ of the giant component, $\cal G'$, versus the fraction of edges deleted, $m$, using different types of edge deletion strategies before first node disconnection in (a) E.U. \& (b) U.S. power grid network (PGN), and, (c) {\it E. coli} \& (d) {\it S. cerevisiae} PPIN. ${\cal D}_0$ is the diameter of the original network. The increase in $\cal D_S $ versus $m$ is initially higher for ${\cal B}_e$ but is eventually highest for ${\cal P}_e$. $\cal L_G$-versus-$m$ curves show similar behavior.} 
\label{Fig.APL.Power}  
\end{figure}

We test these strategies for PGNs of the European Union~\cite{eu-power} and United States~\cite{us-power} by recording change in $\cal L_G$ and $\cal D$ till the first node disconnection. As shown in Fig.~\ref{Fig.APL.Power}, deletion of edges by ${\cal B}_e$ has a strong effect on the increase in $\cal D$ (and $\cal L_G$), initially. Random edge deletion does not lead a significant increase in $\cal D$ (and $\cal L_G$) of the PGNs. Similarly, targeting by $max({\cal K}_e)$ does not lead to a significant increase in $\cal D$ (and $\cal L_G$), at least for the U.S. PGN. The most striking increase in $\cal D$ (and $\cal L_G$) is, however, seen for successive deletions using  $max ({\cal P}_e)$. Thus ${\cal P}_e$ identifies specific edges whose existence is crucial for the network. Damage to these edges affects $\eta$, $\cal L_G$, and $\cal D$ significantly in PGNs.

We also calculate the {\em efficiency}, $\eta$, which is the average of the inverse of all shortest path lengths between any pair of nodes in $\cal G$~\cite{Vrago-effi}: 

\begin{equation}
\eta = \frac{1}{{\cal N}({\cal N}-1)} \sum_{s,t \in {\cal V}; ~ s \ne t} \frac{1}{d(s, t)}
\label{Eq:Efficiency}
\end{equation}

\begin{figure}[htb]%
\vspace{-0.15in}%
\includegraphics[width=0.5\columnwidth, height=5cm]{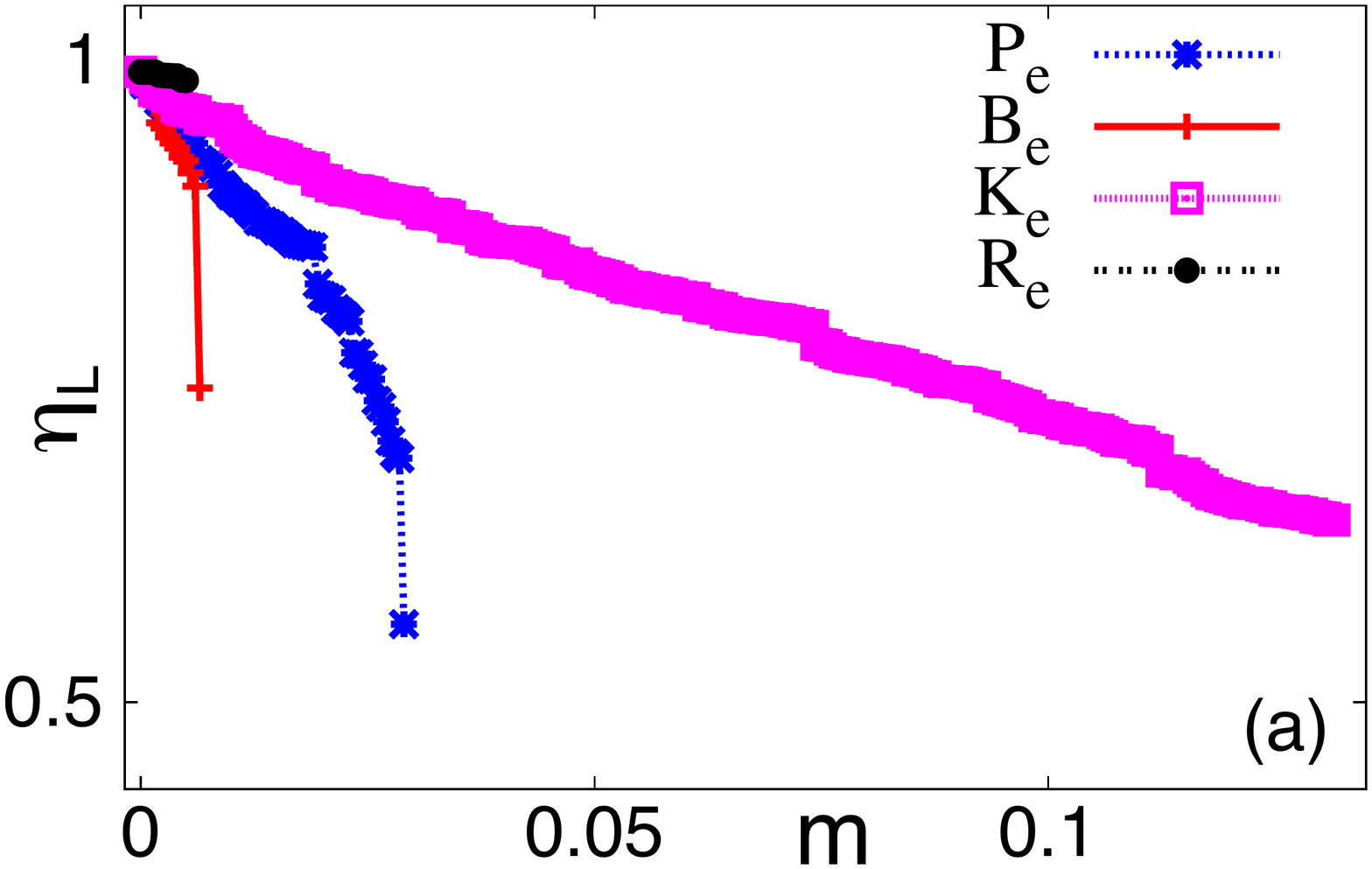}%
\includegraphics[width=0.5\columnwidth, height=5cm]{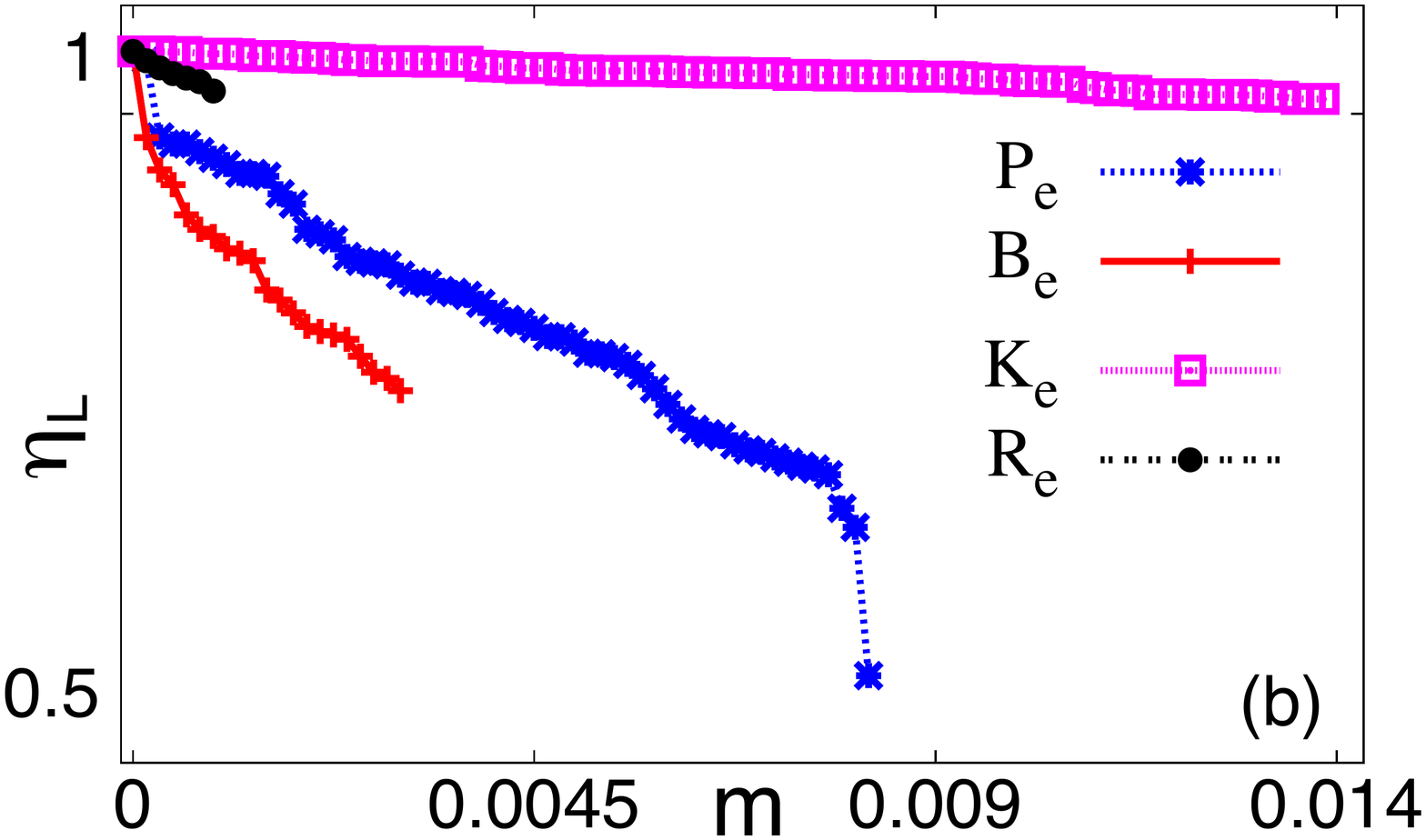}%
\vspace{-0.35in}%
\caption{Loss of efficiency $\eta_{L} = \eta/\eta_0$ versus fraction of edges deleted, $m$, for different strategies  of  (a) EU and (b) US power grid networks, where $\eta_0$ is the efficiency of the original network.}
\label{Fig:Eff}  
\end{figure}

Both E.U. and U.S. PGNs become disconnected by a small number of random edge deletions with insignificant loss of $\eta$, as shown in Fig.~\ref{Fig:Eff}. Notably, connectedness is still maintained for both E.U. and U.S. PGNs when targeted by $max({\cal K}_e)$. However, maximum loss of \textbf{$\eta$} (and increase in $\cal D$ and $\cal L_G$) is observed when edges are targeted by $max ({\cal P}_e)$. Of course, loss of \textbf{$\eta$} in the E.U. PGN is comparable for $max({\cal K}_e)$ and $max({\cal P}_e)$ strategies. However, it is not comparable for the U.S. PGN. Thus, in general, $max({\cal P}_e)$ edges could be very different from $max({\cal K}_e)$ edges. 

\begin{figure*}[tb]%
\vspace{-1.2in}%
\hspace{-0.25in}
\includegraphics[width=0.71\columnwidth, height=10.5cm, width = 6.5cm]{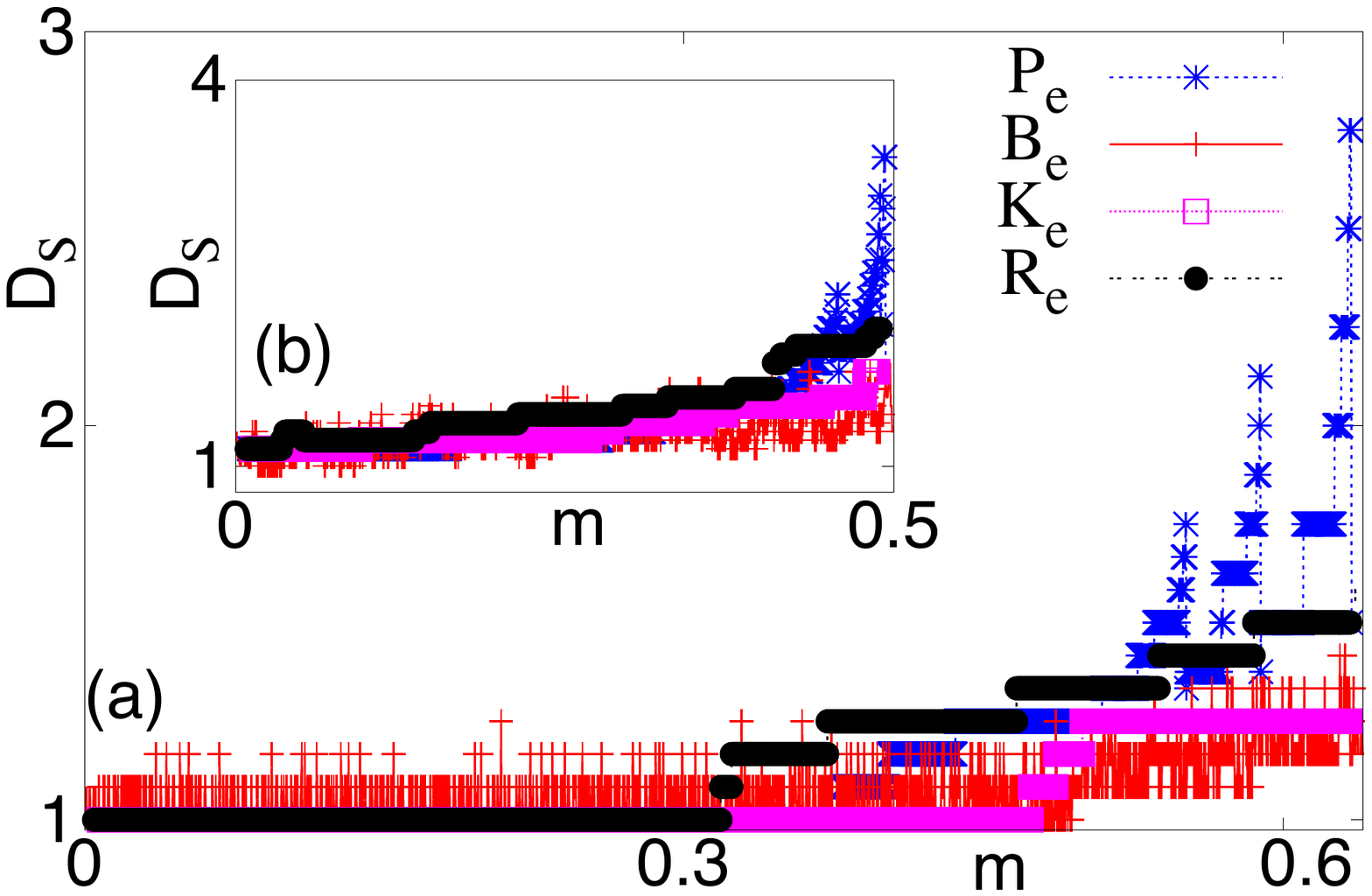}%
\hspace{-0.25in}
\includegraphics[width=0.71\columnwidth, height=10.5cm, width = 6.5cm]{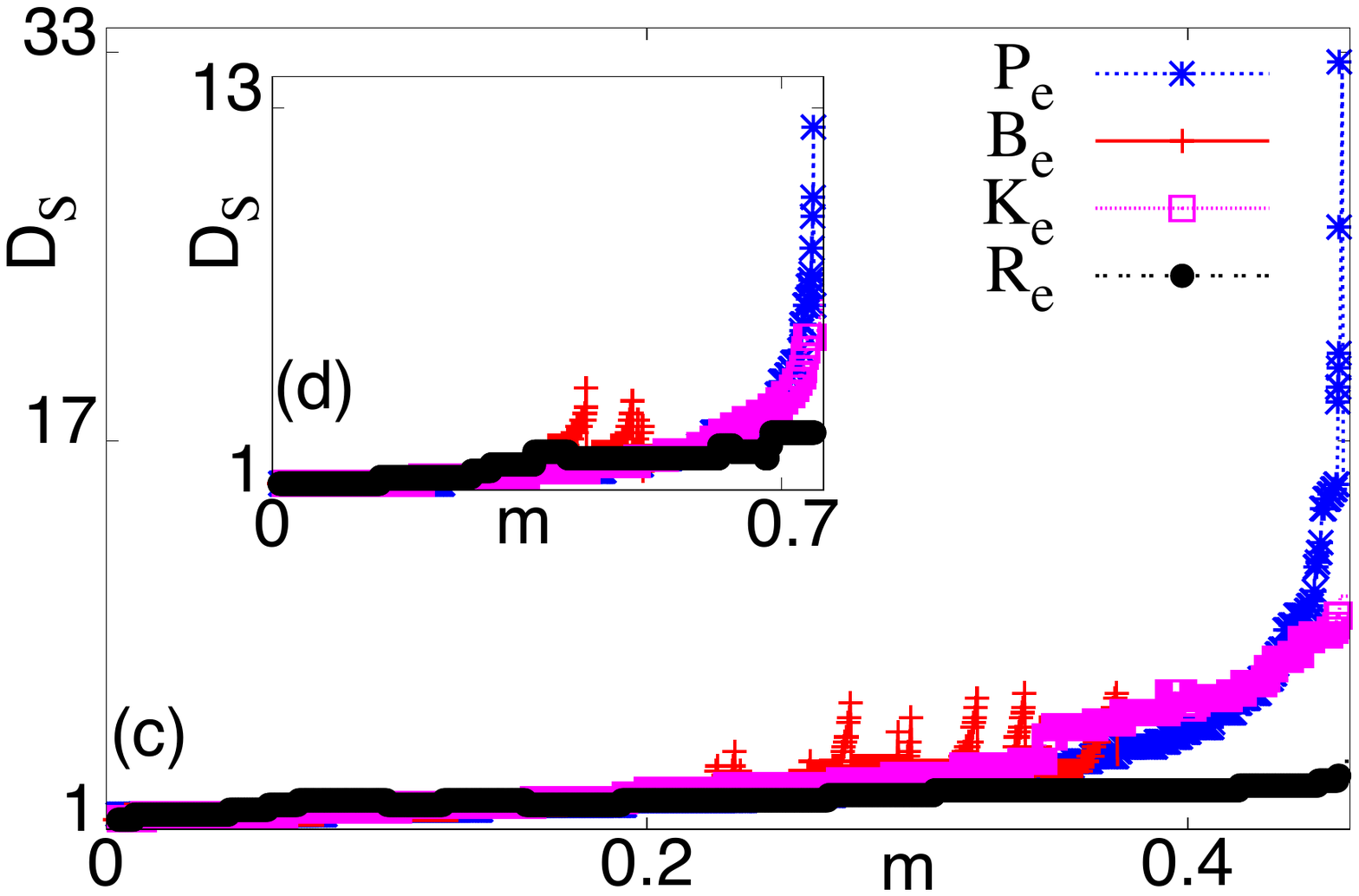}%
\hspace{-0.25in}
\includegraphics[width=0.71\columnwidth, height=10.5cm, width = 6.5cm]{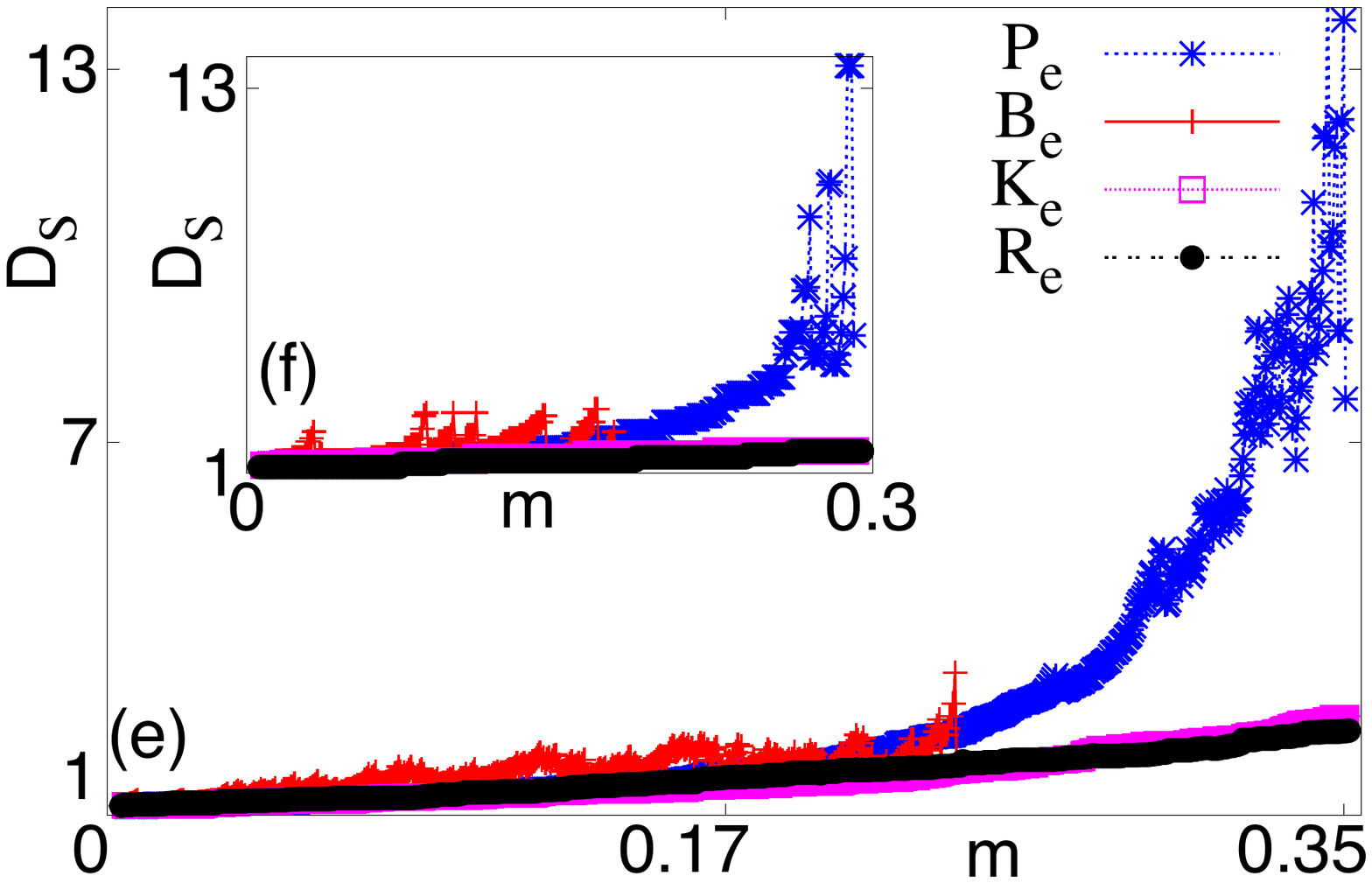}%
\vspace{-1.3in}%
\caption{${\cal D_S}$ of  giant component, $\cal G'$, versus $m$ for (a, b) ER~\cite{er}, (c, d) BA~\cite{ba}, and (e, f) SW~\cite{sw} till $\cal G'$ has only $30\%$ of the original edges of $\cal G$.  (a, c, e) ${\cal N}=1000$. (b, d, f) ${\cal N}=500$.  Data were averaged over $50$ and  $20$ realizations for ${\cal N}=500$ and ${\cal N}=1000$, respectively. Standard  error is negligible. For ER the probability of edge creation is $0.02$. For BA, $m_0 =2$ in (c) and $m_0=3$ in (d). For SW, initially every node has five and four neighbors, respectively, and the probability of edge rewiring is $0.3$ and $0.2$, respectively, in (e) and (f).  ${\cal D_S}$ increases with clustering. Therefore, ${\cal P}_e$ would be important in most real-world networks.}
\label{Fig:ER-SW-BA}
\end{figure*}

We keep deleting edges even after the first node disconnection, until $\cal G'$ contains only $30\%$ of the edges of $\cal G$, as shown in Fig.~\ref{Fig:ER-SW-BA} for BA, ER and SW networks.  Further edge deletions are not conducted because in very small graphs ${\cal P}_e$ loses its meaning, as every edge is usually quite close to most other edges.  
${\cal D_S}$ fluctuates most in ER networks due to the larger number of disconnections. 
Effects due to ${\cal P}_e$ are higher in networks with higher clustering.  Thus  ${\cal P}_e$ should be important for many real-world networks.

\begin{figure}[tb]%
\vspace{-0.2in}%
\hspace{-0.25in}%
\includegraphics[width=1.07\columnwidth, height=5.7cm]{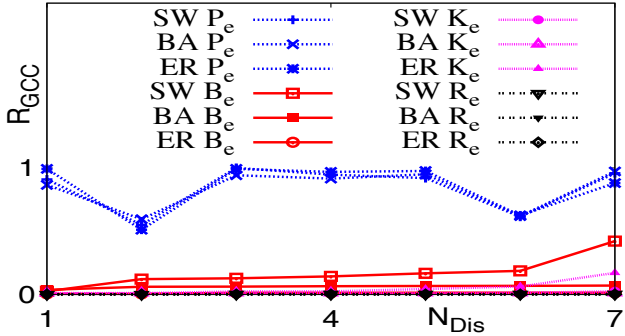}%
\vspace{-0.3in}%
\caption{Ratio of the size of the two biggest components, ${\cal R}_{GCC}$, of $\cal G$ after each disconnection in BA, ER, and SW networks. ${\cal N}_{Dis}$ is the number of disconnections till $\cal G'$ has only $30\%$ of the edges of $\cal G$. Values are the same as those for ${\cal N} = 500$ in Fig.~\ref{Fig:ER-SW-BA}.} 
\label{Fig:GCC-R}
\end{figure}

Great harm can be caused to  a network if it can be broken into two nearly equal parts at each disconnection rather than having a small part disconnected from it. As demonstrated in Figure.~\ref{Fig:GCC-R},  for BA, ER, and SW networks, targeting by ${\cal P}_e$ can be a remarkably successful way to consistently achieve this outcome. 


For all biological networks studied here, we scrutinise the biological significance of edges with the highest ${\cal P}_e$ and ${\cal B}_e$. For this purpose, we calculate
\begin{equation}
{\cal Z(Q}) = \frac{{\cal Q} - \mu({\cal Q})}{ \sigma({{\cal Q}})},  {\cal Q} \in \{{\cal P}_e,  {\cal B}_e\}. 
\label{Eq:Z-score}
\end{equation}
$\mu({\cal Q})$ is the mean and, $\sigma({{\cal Q}})$, the standard deviation of the ${\cal Q}$ distribution. For consistency, we restrict ourselves to the top $2\%$ of edges with ${\cal Z(Q}) \ge 1$.  

First, we observe the effect of edge deletion in $\cal G'$
of yeast
~\cite{Jeong} and {\it E. coli}~\cite{Peregr-Alvarez} PPINs till first disconnection.  As shown in Figs.~\ref{Fig.APL.Power} (c) and (d), removal of interactions by ${\cal P}_e$ increases $\cal L_G$ and $D$  of PPINs the most.  In the {\em E. coli} PPIN, an essential protein, $60kDa$ chaperonin is consistently present at one end of ${\cal Z(P}_e)\ge1$ edges.  This 60kDa protein exhibits a specific stress dependent co-expression with its connected proteins via these ${\cal Z(P}_e)\ge1$  edges~\cite{Viitanen}, alike date hubs~\cite{Han}.

Tropic interactions and energy flow directions in food web networks (FWNs) are represented by directed edges from prey to predators~\cite{Dunne}. The cascade of extinction and role of keystone species are well documented depending on the species- or node-based approach of FWNs~\cite{Allesina,Mills}. Tropic interactions or the edge-based approach might be beneficial for identifying important interactions in the FWN. Herein, we analyze three coral reef FWNs of Cayman Island, Cuba and Jamaica~\cite{Roopnarine} and four FWNs in South Florida ecosystems~\cite{Ulanowicz,Vladimir}.  ${\cal B}_e$ and  ${\cal P}_e$ identify different sets of edges for these FWNs. The root interactions of the dominator tree of corresponding FWNs~\cite{Allesina,Mones} are considered to be important interactions of primary consumers with producers or interactions with the outside environment. These are located at the initial positions in long food chains.  Removal of these interactions may lead to secondary extinction of many species~\cite{Allesina} or may stop the sending of input energy into the FWN from the outer environment~\cite{Ulanowicz,Vladimir}. ${\cal P}_e$ seems to identify these root interactions correctly. ${\cal Z(P}_e)\ge1$ edges include such interactions as primary consumers interacting with planktonic bacteria, phytoplanton, or macrophytes and the environmental input to primary producers and epiphytes. ${\cal B}_e$ identifies other interactions between keystone species such as {\it Diadema}, bivalves, vertebrate detritus, meso-invertebrates, etc~\cite{Bielmyer,Girvan-bet,Sammarco,Dame1,Rahman,Mauchline,Dame2}, successfully. Thus, many food chains possess ${\cal Z(B}_e)\ge1$ edges and their removal might hamper many tropic interactions~\cite{Bielmyer}. 

There has been significant research on detecting emergent behavioral patterns from networks of interconnected neurons. Functional and structural aspects of neural networks are rather well studied in the case of the {\it C. elegans} network. The connectivity data have been obtained from reconstruction of electron microscopy~\cite{White}.  To detect functionally and structurally important synapses, we analyze the network using ${\cal P}_e$ and ${\cal B}_e$.   ${\cal Z(B}_e)\ge1$ identifies RME, AIB, RIA, RIF, AIM, and AEV synapses. These are ring motors and interneurons associated with thermotaxis and backward movement~\cite{Riddle,Wakabayashi, Ohnishi}. However,  ${\cal Z(P}_e)\ge1$ identifies various synapses of AVEL which are solely associated with backward movement of {\it C. elegans}~\cite{Riddle}. 

We also study the brain network of monkeys formed from connectivity data on macaque brain (CoCoMac data sets), where neural fibers connecting different portions of the brain are represented by directed edges~\cite{Stephan}.  Analyzing all neural connections by  ${\cal B}_e$ and  ${\cal P}_e$, we find that two rather different {\em types of edges} in hierarchical information processing pathways are identified by these two metrics.   ${\cal Z(B}_e)\ge1$ identifies interactions which are essentially localized in the intermediate regions of the brain like the prefrontal cortex~\cite{Modha}. Interestingly,  ${\cal Z(P}_e)\ge1$ corresponds to various connections from cortex to thalamus, frontal lobe and temporal lobe, which are the starting interactions of longer information processing pathways from the cortex to other regions of the brain~\cite{Modha}.

\begin{table}[tb]%
\begin{tabular}{|p{3cm}|p{2.25cm}|p{2.25cm}|}
\hline
Edge&${\cal P}_e$&${\cal B}_e$\\
\hline
(1,2) ${\in\cal E}_1^{'}$&0.3472&0.08333\\
\hline
(1,5) ${\in\cal E}_1^{'}$&0.3461&0.08333\\
\hline
(5,6) ${\in\cal E}_1^{'}$&0.2857&0.1167\\
\hline
(2,6) ${\in\cal E}_2^{'}$&0.2812&0.15\\
\hline
(3,2) ${\in\cal E}_2^{'}$&0.2249&0.1667\\
\hline
(6,3) ${\in\cal E}_2^{'}$&0.1875&0.1333\\
\hline
(5,4) &0.0&0.0667\\
\hline
(2,3) &0.0&0.0667\\
\hline
(6,4)&0.0&0.1\\
\hline
\end{tabular}
\caption{${\cal P}_e$ helps in identifying driven edges of ${\cal G(V,E)}$ in Fig.~\ref{Fig:toy}. ${\cal E'}$ and $\{{\cal E}_f\}$  denote all possible sets of  driven edges and edge sets of feedback loops respectively. We observe that  $e\in{\cal E}_i^{'}$ for some ${\cal E}_i^{'}\in{\cal E'}$, $i\in{\mathbb Z_+}$ if ${\cal P}_e>0$ and  if ${\cal E}_i^{'}\notin\{{\cal E}_f\}$. But ${\cal E}_2^{'}\notin{\cal E'}$, because, ${\cal E}_2^{'}\in\{{\cal E}_f\}$ and is self-controllable.}
\label{edge-control}
\end{table}

Finally, under SBD of edge controllability, each node is conceived as acting similarly to a small switchboard-like device~\cite{Nepusz-control}. Nodes map the input signals of the inbound edges to the outbound edges. Figure.~\ref{Fig:toy} depicts an example akin to Ref.~\cite{Nepusz-control}. The maximum matching algorithm is used for the line graph, $\cal L(G)$, constructed from the original network $\cal G = G (V,E)$ for identifying all possible sets of driven edges, ${\cal E}^{'} = \{ {\cal E}_i^{' } : i \in{\mathbb Z_+}\}$, in $\cal G$ under SBD; ${\cal E}_i^{'} \subseteq {\cal E}$. We calculate ${\cal P}_e$ and ${\cal B}_e$,  $e \in {\cal E}$ in Fig.~\ref{Fig:toy}. 
Intuitively,  edges with higher  ${\cal P}_e$ could be driven edges for edge controllability under SBD.  This is consistent with Table~\ref{edge-control} where ${\cal P}_e>0$ for ${\cal E}_1^{'}\in{\cal E'}$.  ${\cal E}_2^{'}= \{(2,6) , (3,2), (6,3)\}$ 
also shows ${\cal P}_e>0$, thus raising the question whether ${\cal E}_2^{'}\in {\cal E'}$?  However, ${\cal E}_2^{'}\in {\cal E}_f$, where $ \{{\cal E}_f\}$ denotes all sets of edges participating in {\em feedback loops} of ${\cal G}$. Thus, ${\cal E}_i^{'} \in \{{\cal E}_f\}, i \in {\mathbb Z_{+}}\Longrightarrow {\cal E}_i^{'} \notin {\cal E'}$,  because if ${\cal E}_i^{'} \in \{{\cal E}_f\}$, ${\cal E}_i^{'}$ is self-controllable.
Again either of $(5,4)$, $(2,3)$, or $(6,4)$ is not a good driven edge because each of these edges can {\em only} control itself. We observe that ${\cal P}_e =0$ for them. This proof-of-concept example illustrates the potential utility of ${\cal P}_e$ to act as an index of edge control centrality of individual edges under SBD.

Even though infrastructure and biological networks have been examined in-depth in this paper,  ${\cal P}_e$ would play a prominent  role in most real-world and random networks, especially large ones. For very small networks, ${\cal P}_e$ becomes irrelevant because almost all edges are then mutually close. The slow poisoning effect due to ${\cal P}_e$ increases with clustering in undirected graphs. Therefore, ${\cal P}_e$ would be important in a wide variety of real-world networks. As with almost any other network metric, its importance would be rather limited in very dense graphs. 

S.J.B. and S.S. thank CSIR, India and UGC, India respectively for financial support.

\end{document}